\def\year{2021}\relax
\documentclass[letterpaper]{article} % DO NOT CHANGE THIS
\usepackage{aaai21}  % DO NOT CHANGE THIS
\usepackage{times}  % DO NOT CHANGE THIS
\usepackage{helvet} % DO NOT CHANGE THIS
\usepackage{courier}  % DO NOT CHANGE THIS
\usepackage[hyphens]{url}  % DO NOT CHANGE THIS
\usepackage{graphicx} % DO NOT CHANGE THIS
\usepackage{natbib}  % DO NOT CHANGE THIS AND DO NOT ADD ANY OPTIONS TO IT
\usepackage{caption} % DO NOT CHANGE THIS AND DO NOT ADD ANY OPTIONS TO IT
\usepackage{xcolor}
\usepackage{amsmath}
\usepackage{breqn}
\usepackage[cal=cm]{mathalfa}
\usepackage{mathtools}
\usepackage{bbm}
\usepackage{multirow}
\usepackage{booktabs}

\usepackage[linesnumbered,ruled]{algorithm2e}
\usepackage[noend]{algpseudocode}
\usepackage[normalem]{ulem}
\makeatletter

\usepackage{amsthm}
\newtheorem{theorem}{Theorem}
\newtheorem{corollary}{Corollary}[theorem]
\newtheorem{lemma}[theorem]{Lemma}
\newtheorem{definition}[theorem]{Definition}

\mathchardef\hy="2D

\DeclarePairedDelimiter\ceil{\lceil}{\rceil}
\DeclarePairedDelimiter\floor{\lfloor}{\rfloor}
\newcommand{\x}{$\times$}

\graphicspath{ {fig/} }

\title{On Provable Backdoor Defense in Collaborative Learning}
\author {
        Ximing Qiao, Yuhua Bai, Siping Hu, Ang Li, Yiran Chen, Hai Li\\
}
\affiliations {
    ECE Department, Duke University \\
    {ximing.qiao, yuhua.bai, siping.hu, ang.li630, yiran.chen, hai.li}@duke.edu
}

\begin{document}

\maketitle

\begin{abstract}
% As collaborative learning allows joint training of a model using multiple sources of data, security problem has been a central concern. Malicious users can upload poisoned data to prevent the model's convergence or inject hidden backdoors. The so called backdoor attacks are especially difficult to detect since the model behaves normally on standard test data but gives wrong output when triggered by certain backdoor keys. Although Byzantine-tolerant training algorithms provide convergence guarantee, provable defense against backdoor attacks remains largely unsolved. A few recent attempts based on randomized smoothing can correct a small number of corrupted pixels or labels but are not directly applicable to collaborative learning. We develop a novel defense based on subset aggregation where we train multiple sub-models on different subsets of data and aggregate their outputs to provide the robust final result. We show that the optimal subset selection can be viewed as a code design problem and derive the upper bound of data utilization ratio. Experiments on non-IID versions of MNIST and CIFAR-10 show that our method with optimal code design significantly outperforms the majority vote baseline.
As collaborative learning allows joint training of a model using multiple sources of data, the security problem has been a central concern.
Malicious users can upload poisoned data to prevent the model's convergence or inject hidden backdoors.
The so-called backdoor attacks are especially difficult to detect since the model behaves normally on standard test data but gives wrong outputs when triggered by certain backdoor keys.
Although Byzantine-tolerant training algorithms provide convergence guarantee, provable defense against backdoor attacks remains largely unsolved.
Methods based on randomized smoothing can only correct a small number of corrupted pixels or labels; methods based on subset aggregation cause a severe drop in classification accuracy due to low data utilization.
We propose a novel framework that generalizes existing subset aggregation methods.
The framework shows that the subset selection process, a deciding factor for subset aggregation methods, can be viewed as a code design problem. 
We derive the theoretical bound of data utilization ratio and provide optimal code construction.
Experiments on non-IID versions of MNIST and CIFAR-10 show that our method with optimal codes significantly outperforms baselines using non-overlapping partition and random selection.
Additionally, integration with existing coding theory results shows that special codes can track the location of the attackers.
Such capability provides new countermeasures to backdoor attacks.
\end{abstract}

\section{Introduction}

As data acquisition increasingly becomes the bottleneck of modern machine learning, collaborative learning on crowd-sourced data gains importance~\cite{vaughan2017making}.
Unfortunately, the distributed nature of collaborative learning opens up an immense attack surface. The system is highly vulnerable to data poisoning attacks~\cite{munoz2017towards} and model poisoning attacks~\cite{baruch2019little}.
Depending on the training protocol, a few malicious users, usually referred to as Byzantine workers, can prevent the model's convergence or introduce hidden backdoor by uploading false data or false gradients.
Extensive research has been conducted on developing Byzantine-tolerant training algorithms~\cite{mhamdi2018hidden, blanchard2017machine, yin2018byzantine}, which target on learning the correct functionally even when a number of users are corrupted.
The central idea around these methods is that false gradients are statistically different from normal gradients, and the false gradients can be detected and removed by an outlier detection process.
Under sufficient assumptions such as smoothness and bounded variance, convergence of the model can be guaranteed.

The problem that Byzantine-tolerant training algorithms cannot solve is the backdoor problem~\cite{li2020backdoor}.
The backdoor attack is a hybrid of training-stage attack and inference-stage attack, which is very different from pure training-stage attacks such as the convergence-preventing attacks, or pure inference-stage attacks such as adversarial examples~\cite{goodfellow2014explaining}.
On one hand, backdoor attacks preserve the model's normal functionality on benign data so that the backdoor is hidden.
On the other hand, a set of attacker-specified triggers (a.k.a. backdoor triggers), such as small patches with particular pixel patterns, can control the model's behavior when those triggers appear in the model's input.
Injecting a backdoor can be viewed as a multitask learning, where the main task is the correct classification on benign data, and a side task is the attacker-controlled behavior on data with backdoor triggers.
Even if Byzantine-tolerant training algorithms can guarantee the convergence, learning of the side task cannot be prevented.
Attacks that successfully bypass Byzantine defenses are shown in various literature~\cite{baruch2019little, bagdasaryan2018backdoor}.

\begin{figure*}[t]
    \centering
    \includegraphics[width=0.85\textwidth]{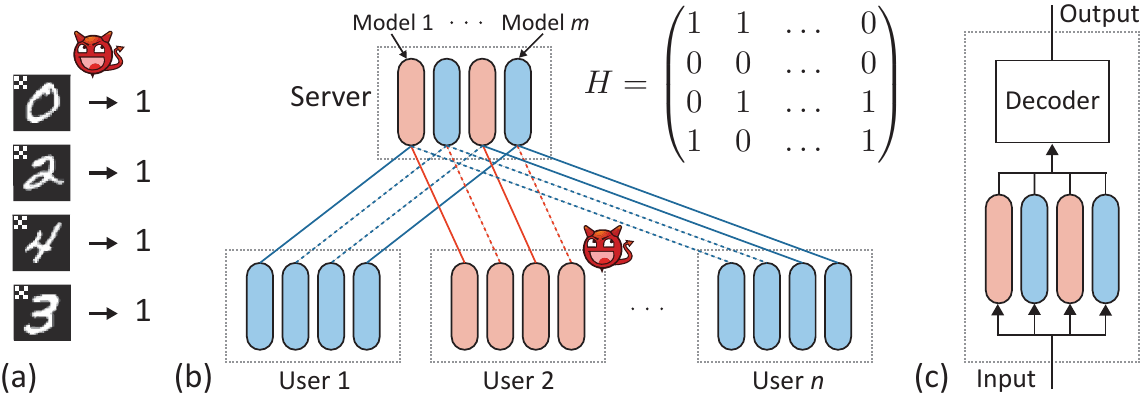}
    \caption{(a) The attacker poisons local data/model to identify the trigger as label ``1''. (b) During training, the model only aggregates data/gradients from the solid lines, so that a few attackers cannot affect all models. (c) During inference, a decoder identifies the correct output from $m$ potentially erroneous outputs.}
    \label{fig:1}
\end{figure*}

Outside the scope of collaborative learning, defense against backdoor attacks is widely studied but mostly based on heuristics.
Typical strategies include rejecting data that produce abnormal activations~\cite{tran2018spectral}, reverse engineering backdoor triggers~\cite{wang2019neural}, masking input regions to remove backdoor triggers~\cite{chou2018sentinet}, and etc.
These defenses are effective against several classical forms of backdoor attacks, but not against some newer variants~\cite{tan2019bypassing, salem2020dynamic}.
Very recently, provable backdoor defenses have been introduced to provide certified robustness against a certain degree of data poisoning, regardless of the exact attack method.
In \cite{weber2020rab} and \cite{wang2020certifying}, the authors extend the randomized smoothing method, a popular algorithm that provides certified robustness against adversarial examples.
Assuming that the attacker's modification to the training dataset is norm limited (e.g., the perturbation to training images is $L_2$ limited or the change of training labels is $L_0$ limited), the defender can introduce noise to the training process and create a ``smoothed'' classifier.
An ensemble model is trained on noisy versions of the training dataset and a smoothed output is produced by a majority vote.
The main drawback of this method is that it requires to train up to $10^3$ models but can only correct a few pixels of modification.
The norm limit is too strict and usually not practical in the context of collaborative learning, where each user controls a large portion of training data and can make arbitrary modifications.

Current state-of-the-art methods are based on subset aggregation/bootstrap aggregation, as in \cite{levine2020DeepPA} and \cite{jia2020intrinsic}.
An ensemble model is trained on \textit{subsets} of the training dataset, and then use majority vote to produce the output.
Selection of the data subsets can be deterministic \cite{levine2020DeepPA} or random \cite{jia2020intrinsic}. 
In the context of collaborative learning, the deterministic method can be interpreted as follows: given $n$ users in total, we can split the users into $m$ groups and train $m$ models on the $m$ groups, then apply majority vote on the $m$ outputs.
The majority vote is robust against $k$ attackers, if the count of the highest vote is $2k+1$ larger than the count of the second-highest vote.
The intuition is that since $k$ attackers can at most change $k$ out of the $m$ outputs, the second-highest vote can never surpass the highest vote.
The random version uses random subsets of users and provides a probabilistic robustness guarantee.
% Although still based on ensembles, these methods are slightly cheaper than the randomized smoothing methods, since each model is trained on a subset of data.
Although subset selection reduces the training cost, it also negates the benefit of collaborative learning, which is to train models on more data.
To defend $k$ attackers, each model can be trained on at most $1/(2k+1)$ portion of the total available data.
In extreme case, users have no collaboration at all and simply use local models to vote for their final result.
% As we will show in experiments, restricting the amount of training data can significantly lower the model's accuracy in the collaborative learning setting.

In this work, we find that simple subset selection methods such as random selection or non-overlapping partition cannot suffice the complicated condition of collaborative learning, especially on non-IID datasets.
Centered around the idea that each model in the ensemble should be trained on as much data as possible, we seek new subset selection algorithms and new aggregation methods.
We show that the optimal subset selection can be interpreted as a code design problem.
As illustrated in Figure~1, training of the ensemble model is characterized by a binary code matrix and the majority vote aggregation is generalized to a decoder.
The new framework allows us to provide a theoretical bound of data utilization ratio and derive the optimal code construction.
The main result is that with additional assumptions of the \textit{independence} and \textit{determinism} of the backdoor attack, we can correctly decode the final result even when the majority of the models give wrong outputs.
This leads to a data utilization ratio beyond the $1/(2k+1)$ bound of majority vote, and results in a significant boost in classification accuracy.
Additionally, the new framework allows us to borrow rich results from coding theory and create novel applications of the defense.
One example is based on superimposed codes~\cite{kautz1964nonrandom} originated from communication and group testing.
The codes allow us to track the location of the attacker (whenever a backdoor is triggered), providing new countermeasures to the backdoor attack.

\section{Problem Formulation}

\subsection{Subset Selection and Code Matrix}
In this section, we build the basic framework and show that subset selection can be formulated as a code design problem.
Here we first introduce the matrix representation of the subset selection.
Assume that we train $m$ model using data from $n$ users and each model is trained on a subset of users.
The selection of users can be represented by a $m\times n$ binary matrix, referred to as a \textit{code} $H$.
This matrix has its element $H_{ij}=1$ when model $i$ accepts the data (or gradients) from user $j$ and $H_{ij}=0$ when it rejects the data (or gradients).

The matrix representation allows us to describe subset aggregation-based defenses in a uniform way.
In \cite{levine2020DeepPA}, their defense splits the $n$ users in to $m$ groups without overlap, and then train $m$ models on those $m$ groups.
The equivalent code is a matrix with one-hot columns (since there is no overlap). It can also be viewed as a diagonal matrix extended by repeating columns.
In \cite{jia2020intrinsic}, their train $m$ models using $m$ independent random subsets of users (with possible overlaps). The equivalent code is a random matrix with a constant number of 1's per row.
Examples of the codes are given in Figure~\ref{fig:2}(a) and 2(b) for $n=6$.

\subsection{Defense Objectives and Notation}
With the matrix representation, we can specify objectives of defense by properties of code matrices.
In this work, we are interested in the three types of defenses listed below.
% A code matrix $H$ has a row weight $r$ if the number of 1's in each row of $H$ is greater than or equal to $r$.
Assuming that there are less than or equal to $k$ attackers among the $n$ users, we can have:
\begin{itemize}
    \item \textit{Backdoor Detection Codes} (BDC) that detect whether an attack happens without necessarily giving the true label;
    \item \textit{Backdoor Correction Codes} (BCC) that detect the attack and recover the true label;
    \item \textit{Backdoor Tracking Codes} (BTC) that detect the attack, recover the true label, and identify the location of the attackers.
\end{itemize}
To define the codes more specifically, we introduce $r$ as \textit{row weight}, the number of 1's in a row.
We say a code matrix $H$ has a row weight $r$ if the number of 1's in every row of $H$ is greater than or equal to $r$.
The row weight is an important characteristic of a code as it represents the amount of data used to train each model.
When each user contribute the same amount of data, the \textit{data utilization ratio} can be defined as $r/n$.
For a given set of $k$, $r$, and $n$, we search for codes that satisfy the above properties with the least number of rows, as the row number decides the cost of storing and running the ensemble model.
The three types of codes are denoted as BDC$(k,r,n)$, BCC$(k,r,n)$, and BTC$(k,r,n)$ respectively.
% BDC codes are not practically useful but have strong theoretical relations with BRC codes, which will be discussed later.
% In this section, we focus on codes with $k=1$.
% Apparently, a BDC code should not contain any column equals to $\mathbf{1}$.
% The code in Figure~2(a) is BDC$(1,3,7)$ but not BRC.
% The code in Figure~2(b) is is not any of the BDC, BRC, or BTC.

\section{A Toy Example: Defending a Single Attacker}

\subsection{Binary Classification}
We start with the simple case of defending a single attacker.
Given a code matrix $H$, we describe the attacker as a one-hot vector $\mathbf{x}$ and the models' outputs as a vector $\mathbf{y}$ (all vectors are by default column vectors).
The attack vector has $x_j=1$ if user $j$ is the attacker.
Here we introduce our first \textit{independence} assumption, stating that the injection of a backdoor should be independent to data from those users besides the attacker.
If the backdoor can be successfully injected when a model is trained on all users, the same backdoor should be injected when the model is trained on a subset of users (as long as the subset includes the attacker).
Under an independent attack, $\mathbf{y}$ has two possible outcomes:
If the true label is $0$ and the attack target is $1$, then $y_i=1$ if $\exists j, H_{ij}=1 \land x_j=1$, or in short, $\mathbf{y}=H\mathbf{x}$.
If the true label is $1$ and the attack target is $0$, then $\mathbf{y}=1-H\mathbf{x}$.

\begin{figure}[tb]
    \centering
    \includegraphics[width=\columnwidth]{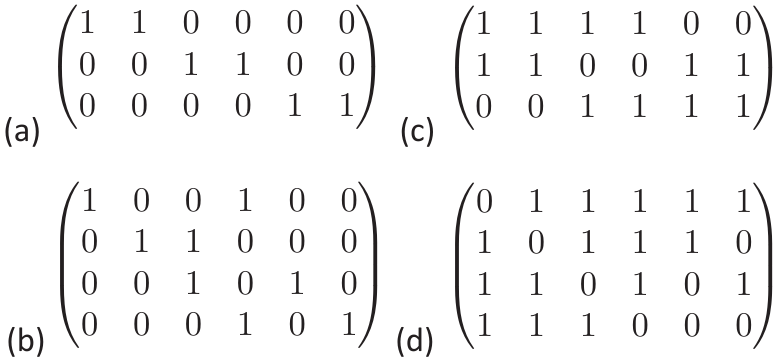}
    \caption{Subset selection in matrix form.}
    \label{fig:2}
\end{figure}

For BDC codes to detect the attack, $H$ must not contain any column that equals to $\mathbf{1}$ (a vector of all 1's) so that at least one model is not backdoored.
Meanwhile, $H$ should not include any zero columns or zero rows, otherwise a user's data are not used or a model is not trained.
For BCC codes to find the true label, the code additionally cannot include two complementary columns, or $\forall i\forall j, H_{\cdot i}\oplus H_{\cdot j}\neq\mathbf{1}$, in which $\oplus$ represents the XOR operation.
For BTC codes to track the location of the attacker (i.e., reconstructing $\mathbf{x}$ from $H$ and $\mathbf{y}$), we need to further ensure that $H$ does not have two identical columns.
When an attack happens, the attacker can be uniquely identified by matching $\mathbf{y}$ or $1-\mathbf{y}$ to one of the columns of $H$.
Examples of the BCC and BTC codes are given in Figure~\ref{fig:2} (c) and (d).
Note that although codes (a) and (c) are both valid BCC codes with the same number of rows, code (c) has a higher row weight $r$, meaning that each model is trained on more data.
However, code (c) cannot be decoded by majority vote and requires more complicated methods.
The random matrix given in code (b) is BCC.
In general, a random matrix is BCC by a probability approaches 1 when $m$ or $n/r$ is large enough.

\subsection{Multi-Class Classification}
In a multi-class classification scenario, the class prediction $\mathbf{y}$ is no longer a binary vector.
To make the above codes available, we make a further assumption of the \textit{determinism} of the attack.
An attack is deterministic if the same input image with the same  backdoor trigger is always classified to the same class.
In other words, the attack target is deterministically decided by the attacker and will not be altered by the randomness during training.
When clean models (non-backdoored models) are accurate enough, the class predictions should contain only two labels: one being the attack target and the other being the true label.
As such, we can map one label to 1 and the other label to 0 so that the problem is equivalent to the binary case.
Problems with inaccurate clean models with be discuss later.

\subsection{The Number of Attackers}
In all the above discussions, $k$ refers to the number of \textit{cooperative} attackers that use the \textit{same} backdoor trigger.
For the case that many independent groups of attackers attack the model using different triggers, $k$ is decided by the size of the largest group.
For example, the above codes (a) to (d) are effective codes when all users are attackers, as long as the attackers are independent and use different triggers.
If any two users share the same trigger, we then need more powerful codes with $k=2$, which are introduced in the next section.

\section{Defending Multiple Attackers}

In this section, we extend the defense to multiple attackers $k>1$ and derive the construction of optimal BDC, BCC, and BTC codes.
Here we begin with the formal definition of codes with arbitrary $k$.
For a code matrix $H$, we denote the Boolean sum of any $k$ columns of $H$ as a $k$-sum, and the Boolean sum of any less than or equal to $k$ columns as a $\bar{k}$-sum.
Comparing to the $k=1$ case in Section 3, the following definition replaces the conditions on \textit{single columns} to conditions on $\bar{k}$-sums.
The reason of using $\bar{k}$-sums is due to the independency assumption: a model is backdoored if it is trained on any of the $1\sim k$ attackers.
% The interpretation of the $\bar{k}$-sum is that if a model is trained on any of the $1\sim k$ attackers, the model is backdoored.
\begin{definition}
\textbf{(1)} $H$ is BDC$(k,r,n)$ if $H$ contains no zero column in its $n$ columns, has at least $r$ 1's in each row, and has no $\bar{k}$-sum equals to $\mathbf{1}$.
\textbf{(2)} $H$ is BCC$(k,r,n)$ if $H$ is BDC$(k,r,n)$ and has no XOR of any two $\bar{k}$-sums equals to $\mathbf{1}$.
\textbf{(3)} $H$ is BTC$(k,r,n)$ if $H$ is BCC$(k,r,n)$ and has no two $\bar{k}$-sums equal to each other.
\end{definition}

% The major difference between BCC codes and the classical error correcting or superimposed codes is that BCC includes both XOR operations and OR operations.
% The linearity of XORs and the monotonicity of ORs are no longer useful to analyze BCC codes.
% If we relax the conditions by replacing XORs to ORs (since $A\oplus B=1\Rightarrow A\lor B=1$), it can be proved that $m\ge 2k+1$ (see Appendix).
% In that case, a majority vote with each model trained on mutually exclusive $N/(2k+1)$ users is an optimal code, which is a rather trivial solution.

\subsection{Minimal Backdoor Correction Codes}

Direct construction of BDC and BCC codes of any size can be difficult so we start by looking at a special case with minimal $n$.
The corresponding codes are denoted as minimal codes.
\begin{lemma}
For $k\ge1$, $r\ge1$, the minimal $n$ for BDC$(k,r,n)$ and BCC$(k,r,n)$ to exist is $n=k+r$.
\end{lemma}
\textit{Proof.}
If $n<k+r$ then all rows have $<k$ 0's and all $k$-sums are $\mathbf{1}$. \hfill\qedsymbol

% We are mostly interested in codes with minimal $m$, since codes with larger $m$ can always be constructed by duplicating rows.
% Also in a practical defense, smaller $m$ means fewer models and lower computation/communication costs.
When $n=k+r$, we can give the tight lower bound of $m$ for all BDC$(k,r,k+r)$ codes:
\begin{theorem}
For $k\ge1$, $r\ge1$, if $H$ is BDC$(k,r,k+r)$, then its number of rows $m\ge\binom{k+r}{k}$.
The solution for $m=\binom{k+r}{k}$, denoted as $\mathcal{H}^{(k,r)}$, is unique up to row and column permutations and can be constructed by:
\begin{align}
\mathcal{H}^{(k,1)}&=I_{k+1},\\
\mathcal{H}^{(1,r>1)}&=\begin{bmatrix}
    \mathbf{1} & \mathcal{H}^{(1,r-1)}\\
    0 & \mathbf{1}^T\end{bmatrix},\\
\mathcal{H}^{(k>1,r>1)}&=\begin{bmatrix}
    \mathbf{1} & \mathcal{H}^{(k,r-1)}\\
    \mathbf{0} & \mathcal{H}^{(k-1,r)}\end{bmatrix}.   
\end{align}

\end{theorem}
\textit{Sketch of proof.}
For the lower bound of $m$, first assume each row to have exactly $k$ 0's. For each row, there is only one $k$-sum to produce 0 in this row. 
When rows are unique, each $k$-sum can have only one 0, and the location of this 0 is unique. Since there are $\binom{k+r}{k}$ different $k$-sums, $m$ cannot be smaller than $\binom{k+r}{k}$.
Cases with non-unique rows and rows with $<k$ 0's can be reduced to the above case.
The construction can be verified by induction and the uniqueness is obvious. \hfill\qedsymbol

Next we show that BCC and BDC codes are closely connected.
In fact, a BCC code can always be constructed from a BDC code by adding one extra row of 1's, meaning the cost of backdoor correction is almost negligible if the attack can be detected:
\begin{lemma}
\label{bdc-to-BCC}
If $H$ is BDC$(k,r,n)$, then $\begin{bmatrix}\mathbf{1}^T\\H\end{bmatrix}$ is BCC$(k,r,n)$.
\end{lemma}
\textit{Proof.} 
All $\bar{k}$-sums have 1 in the first row and $1\oplus1=0$. \hfill\qedsymbol

This leads the main result about minimal BCC codes.
In fact, minimal BCC codes and minimal BDC codes are the same when $r>1$.
\begin{theorem}
For $k\ge1$, $r>1$, the minimal BCC$(k,r,k+r)$ has $m=\binom{k+r}{k}$.
For $r=1$, the minimal BCC$(k,1,k+1)$ has $m=k+2$.
\end{theorem}
\textit{Sketch of proof.}
Prove by induction:
$\mathcal{H}^{(1,r>1)}$ has the XOR of any two $\bar{k}$-sums $\neq\mathbf{1}$ since each column has more 1's than 0's.
For $\mathcal{H}^{(k>1,r>1)}$, if one $\bar{k}$-sum includes the first column then its first $\binom{k+r-1}{k}$ rows are all 1's.
The XOR has at least one 0 in these rows since a $\bar{k}$-sum in $\mathcal{P}^{(k,r-1)}$ has at least one 1.
If no $\bar{k}$-sum includes the first column then the problem is reduced to the $\bar{k}$-sums of $\mathcal{H}^{(k-1,r)}$.
For $r=1$, obviously $I_{k+1}$ is BDC but not BCC. Lemma~\ref{bdc-to-BCC} gives $m=k+2$. \hfill\qedsymbol

Figure 3(a) shows some examples of minimal BCC codes to visualize the pattern of recursive construction.
At this point we obtain BCC codes that can defend any number of attackers $k$ with any large data utilization ratio $r/n$ for the special case $n=k+r$.
Next we look at how to construct general BCC codes for larger $n$.

\begin{figure}
    \centering
    \includegraphics[width=0.7\columnwidth]{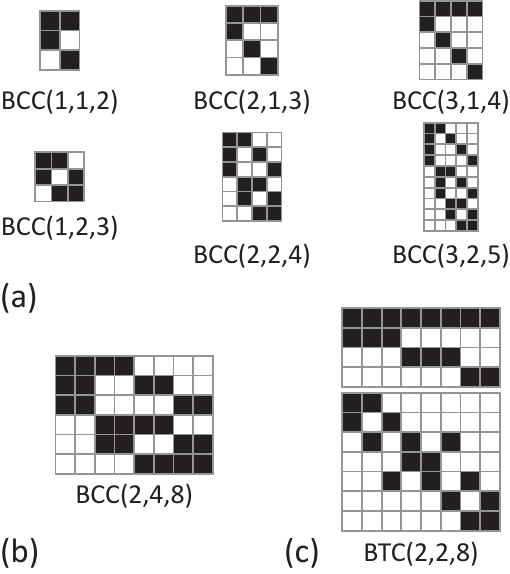}
    \caption{Codes for multi-attacker defense (black represents one and white represents zero).}
    \label{fig:3}
\end{figure}

\subsection{General Backdoor Correction Codes}
The above result can be generalized to BCC codes with any large $n$ by column duplication.
For given $k$, $r$, and $n$, we calculate the relative row weight $r/n$, select a proper BCC$(k,r_0,k+r_0)$ with $r_0/(k+r_0)\approx r/n$ as a starting point, and duplicate its columns until $r$ and $n$ are satisfied:
\begin{corollary}
\label{BCC}
For $k\ge1$, $r\ge1$, $n\ge k+r$, $p=\floor{\frac{n-r}{k}}$, and $r_0=\ceil{\frac{r}{p}}$, let $H$ with columns $\left[\mathbf{h}_1,\mathbf{h}_2,\dots,\mathbf{h}_{k+r_0}\right]$ be BCC$(k,r_0,k+r_0)$, then the following code is BCC$(k,r,n)$:
\[
\bigg[\mathbf{h}_1,\dots,\mathbf{h}_{n-p(k+r_0)},\underbrace{H,\dots,H}_{\text{repeat p times}}\bigg].
\]
\end{corollary}
\textit{Proof.}
For any $\bar{k}$-sum in the above code there exists an equal $\bar{k}$-sum in $H$. Therefore $H$ being BCC implies the above code being BCC. The row weight $\ge pr_0\ge r$. \hfill\qedsymbol

Figure 3(b) gives an example BCC$(2,4,8)$ generated from BCC$(2,2,4)$.
The result obtained by column duplication might not be optimal when $n$ is not divided by $r$.
However, since finding the optimal solution is NP-hard in general, we consider the construction good enough.
% The minimal BCC code is duplicated by 3 times to match the larger $n=12$.
% Such construction gives an upper bound of $m$ for any large $n$.
% In practice, this suggests that \textit{the cost of backdoor resistance is independent of the absolute value of $n$} and mostly determined by the number of attackers $k$ and the relative amount of data $r/n$ used to train each model.
% \textbf{Implications of the result}
The result gives several important implications:
\begin{itemize}
    \item For fixed $k$ and $r/n$, the cost of backdoor correction is independent of the absolute value of $n$.
    \item For fixed $k/n$ and $r$, the cost grows linearly.
    \item For fixed $k/n$ and $r/n$, the cost grows exponentially.
\end{itemize}
Such a result suggests a fundamental tradeoff between data utilization and backdoor robustness.
When either of the data utilization or backdoor robustness is measured by absolute values, the total cost is manageable.
However, if the goal is to achieve both high data utilization ratio and robustness ratio, there is no scalable solution.
% Note that the above result only provides the lower bound of the cost, assuming perfect classification accuracy on clean data.
% In practical defenses, higher cost should be expected.
% At last, we show that this construction is optimal up to a constant factor.
% \begin{theorem}
% For $k\ge1$, $r\ge1$, $n\ge k+r$, if $H$ is BCC$(k,r,n)$, then its number of rows $m\ge???$.
% \end{theorem}
% \textit{Sketch of proof.}
% When $k=2$ the problem of finding minimal BDC$(2,r,n)$ codes is equivalent to finding the minimal $m$ to cover the edges of a complete graph $K_n$ by $m$ cliques of size $\le n-r$.
% This problem was studied in~\cite{orlin1977contentment}.

\subsection{Backdoor Tracking Codes}

% For BTC codes, the attacker tracking capability requires each $\bar{k}$-sum of $H$ to be unique and not covering any other columns not involved in the sum, i.e., $k$-disjunct.
% A code being BTC also implies it being BCC.
% This leads to the following definition:
% \begin{definition}
% \label{def:btc}
% A code $H$ is BTC$(k,m,N)\hy r$ if $H$ is BCC$(k,m,N)\hy r$ and $k$-disjunct.
% \end{definition}

Finally, we show that BCC codes can be concatenated with other familiar codes in coding theory to achieve additional properties and create novel defenses.
For backdoor tracking, BTC codes further require the matrix to have no two $\bar{k}$-sums equal to each other.
% The minimal $m$ for BTC codes clearly depends on $n$ since identifying $k$ attackers requires at least $\log\binom{n}{k}$ bits of information.
In literature about superimposed codes and group testing, a matrix with unique $\bar{k}$-sums is called a $\bar{k}$-separable matrix.
Here we directly construct BTC codes using the existing results of $\bar{k}$-separable matrices:
\begin{corollary}
If $H_1$ is BCC$(k,r,n)$, $H_2$ is $\bar{k}$-separable, and the row weight of $H_2$ is at least $r$, then $H=\begin{bmatrix}H_1\\H_2\end{bmatrix}$ is BTC$(k,r,n)$.
\end{corollary}
\textit{Proof.}
The code $H$ is BCC$(k,r,n)$ since $H_1$ is BCC$(k,r,n)$ and $H_2$ has row weight $r$. The code $H$ is $\bar{k}$-separable since $H_2$ is $\bar{k}$-separable. Therefore $H$ is BTC$(k,r,n)$.  \hfill\qedsymbol

Previous study shows that $\bar{k}$-separable matrices with $n$ columns have a lower bound of $m>O(k\log n)$ and a typical construction of $m=O(k^2\log n)$~\cite{du2000combinatorial}.
% and the solutions are sparse, having a small $r$.
Since BCC codes typically have a much smaller size, concatenating a BCC code with a separable matrix gives a good enough BTC code.
Figure~3(c) shows an example of a BTC$(2,2,8)$ code, with its top 4 rows a BCC$(2,2,8)$ code and bottom 7 rows a $\bar{2}$-separable matrix.
\section{Decoding Algorithm}

In Section 4, we study theoretical properties of the codes in an idealized setting and assume that the class prediction vector $\mathbf{y}$ contains only two groups: the true label and the attack target.
This section fills the gap between the idealized code design and practical defenses, where the noise in $\mathbf{y}$ (i.e. inaccurate classification of clean models) is considered.

Here we propose a probabilistic solution.
We model the noise with a probability distribution and decode the noisy output by probability maximization.
Given a set of validation data, we can statistically estimate the distribution of clean data classification by confusion matrices.
For model $i$, $C^{(i)}_{jk}$ represents the probability of classifying clean data in class $j$ as class $k$.
The attack can be specified by a prior probability of the attack $A$, a success rate of the attack $S$, and a distribution of the number of attackers $Q(\|\mathbf{x}\|_1)$.
Given the class predictions $\mathbf{y}$, the number of users $n$, and the number of classes $c$, the probability of the models being attacked is given by:
\begin{gather*}
    Pr(\text{attack}=\text{True}|\text{pred}=\mathbf{y})=\\
    \frac{A\sum_{\mathbf{x},t,l} p(\mathbf{x},\mathbf{y},t,l)}{A\sum_{\mathbf{x},t,l} p(\mathbf{x},\mathbf{y},t,l)+(1-A)c\sum_l\prod_i C^{(i)}_{ly_i}},\\
    \text{ in which }
    p(\mathbf{x},\mathbf{y},t,l)=\frac{Q(\|\mathbf{x}\|_1)}{\binom{n}{\|\mathbf{x}\|_1}}\times\\
    \prod_i\begin{cases}S\delta_{ty_i}+(1-S)C^{(i)}_{ly_i}&\text{if }\sum_j H_{ij}x_j>0\\C^{(i)}_{ly_i} &\text{otherwise}\end{cases}.
    % \text{ and }
    % \delta_{ij}=\begin{cases}1\text{ if }i=j\\0\text{ otherwise}\end{cases}.
\end{gather*}
Here $\delta_{ij}=1$ if $i=j$ or $0$ if $i\neq j$. 
% The probabilistic model assumes: (1) the attack is deterministic (one trigger only has one fixed target, as described in the $S\delta_{ty_i}$ term of the above equation, in which $t$ is the target), and (2) a failed attack does not affect the model's prediction (as shown in the $(1-S)C^{(i)}_{ly_i}$ term, in which $l$ is the true label). 
The true label $l$ and the attackers $\mathbf{x}$ can be obtained by maximizing over the distributions $Pr(\text{label}=l|\text{pred}=\mathbf{y})\sim A\sum_{\mathbf{x},t} p(\mathbf{x},\mathbf{y},t,l)+(1-A)c\prod_i C^{(i)}_{ly_i}$
and
$Pr(\text{attackers}=\mathbf{x}|\text{attack}=\text{True},\text{pred}=\mathbf{y})\sim \sum_{t,l}p(\mathbf{x},\mathbf{y},t,l).$
See Appendix for detailed derivations.

In the noisy setting, codes with a high $k$ can defend a reduced number of attackers that is less than $k$.
For example, the code BCC$(2,4,8)$ with $k=2$ in Figure~\ref{fig:3}(b) can reliably defend one attacker but not two.
If the six models have five outputs of class A and one output of class B, then there could be two equally likely cases: 1) two attackers have attack target A and the true class is B, and 2) the true class is A and one model misclassifies it as B.
For a single attacker, the attack will results in three outputs of class A and three outputs of class B.
The chance for three models to make the same misclassification is small.

% \subsection{Sharing backbone networks}
% % \textbf{Sharing backbone networks.}
% When computation and communication resources are limited or $m$ is too large, duplicating the whole model for $m$ times becomes unrealistic.
% In such cases, we split the network to a frontend (the backbone network) and a backend (the last few layers), and duplicate only the backend (recall Figure~1(d)).
% The shared frontend aggregates the gradients from all users and the duplicated backends are trained according to the code $H$.
% The key motivation here is that the attack is most effective when the whole network is backdoored end-to-end.
% A model with a backdoored frontend can still produce right predictions when the backend is clean, if the backend is large enough.
% The size of the backend is decided empirically to offer a tradeoff between security and performance.
\section{Experiments}

% In Section 4 we derive the tradeoff between defending more attackers (higher $k$) and training more accurate models (higher $r/n$).
% A code with small $k$ meets unavoidable failures when the number of attackers is above the limit.
% A code with small $r/n$ gives unreliable predictions even when no attack happens.
% With a limited budget of model redundancy (limited $m$), choosing the right code is critical in practical defenses.
% In this section we explore how different codes with different parameter choices perform on different datasets and attack settings.

\subsection{Experiment Setup}
% \textbf{Experiment setup.}
We evaluate the proposed defense on MNIST~\cite{lecun1998gradient} and CIFAR-10~\cite{krizhevsky2009learning}.
The models we tested include a 4-layer CNN for MNIST and ResNet18~\cite{he2016deep} for CIFAR-10.
The attack method is based on direct data poisoning with random 3px\x3px backdoor triggers and random target classes~\cite{qiao2019defending}.
We use 12 users for backdoor correction and 16 users for backdoor detection (for the ease of code construction), with 1$\sim$3 attackers.
Each attacker poisons 5\% of their data to perform the attack.

\begin{figure}
    \centering
    \includegraphics[width=\columnwidth]{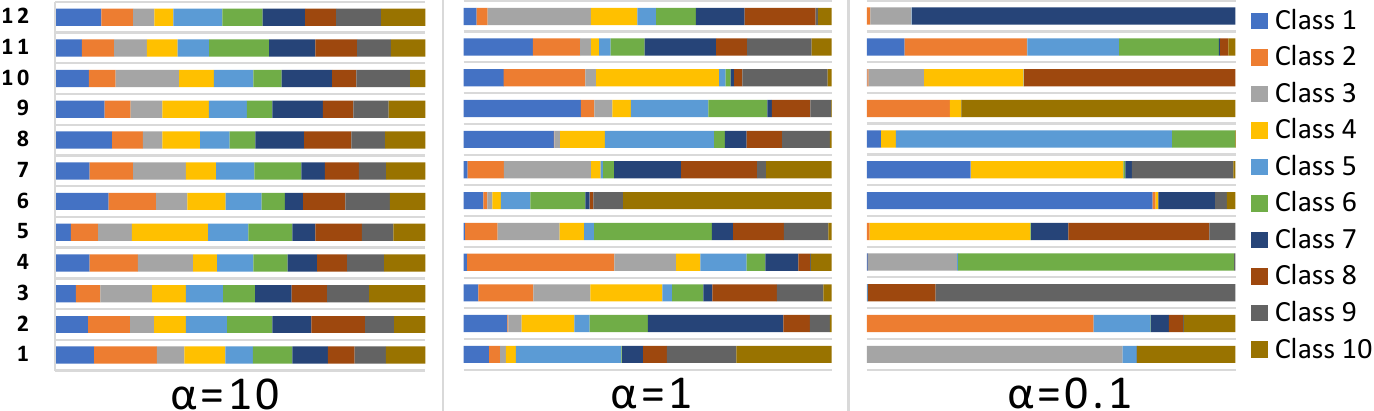}
    \caption{Examples of class distributions controlled by $\alpha$. Each color represents a class and each row represents a user.}
    \label{fig:dirichlet}
\end{figure}

To synthesize the non-IID data, we follow the method from~\cite{hsu2019measuring} and use Dirichlet distribution to decide the number of data per class (a.k.a. class distribution) for each user.
A hyper-parameter $\alpha$ (concentration parameter of Dirichlet distribution) controls the degree of unevenness of the class distribution.
In the experiments we use $\alpha=10$, 1, and 0.1.
Examples of per-user class distributions are given Figure~\ref{fig:dirichlet}.

\begin{figure}
    \centering
    \includegraphics[width=0.8\columnwidth]{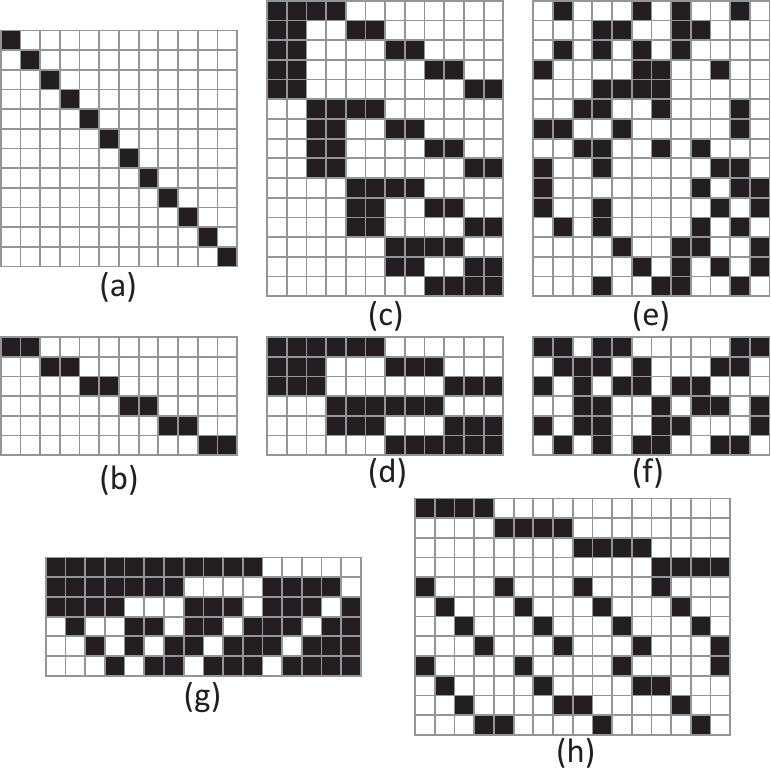}
    \caption{Six BCC codes and two BTC codes used in experiments.}
    \label{fig:4}
\end{figure}

We investigate eight different codes, plotted in Figure~\ref{fig:4}, including six codes for backdoor correction and two codes for backdoor tracking.
Codes (a) and (b) represent non-overlapping partition, following~\cite{levine2020DeepPA}.
Codes (c) and (d) are the proposed optimal BCC$(4,4,12)$ and BCC$(2,6,12)$.
Codes (e) and (f) are random matrices with constant row weight, following~\cite{jia2020intrinsic}.
The row weights are set to 4 and 6 to match with code (c) and (d).
Codes (g) and (h) are backdoor tracking codes in which (g) is BTC$(1,11,16)$ and (h) is BTC$(2,4,16)$.

% \subsection{Evaluation Metric}
% Introduce certified accuracy.

% \subsection{Backdoor detection}
% \textbf{Backdoor detection.}
% We measure the backdoor detection capability of all codes under different number of attackers, and plot the ROC curves in Figure~\ref{fig:roc}.
% The high clean data accuracy on MNIST allows our defense to reach its theoretical limit.
% All codes achieve a near-perfect performance when defending one attacker, code 1 and 6 start to fail when facing two attackers, while code 5 and 7 remain effective when three attackers appear.
% Result from CIFAR gives more insight about the tradeoff between $k$ and $r$.
% While code 5 and 7 with $k=3$ perform the best when defending two or three attackers, they perform worse than other codes when defending one attacker.
% This suggests that the optimal code always depends on the exact attack setting.

% \begin{figure*}
%     \centering
%     \includegraphics[width=\textwidth]{fig/ROC.pdf}
%     \caption{ROC curves of backdoor detection under different number of attackers.}
%     \label{fig:roc}
% \end{figure*}

\begin{figure*}
    \centering
    \includegraphics[width=\textwidth]{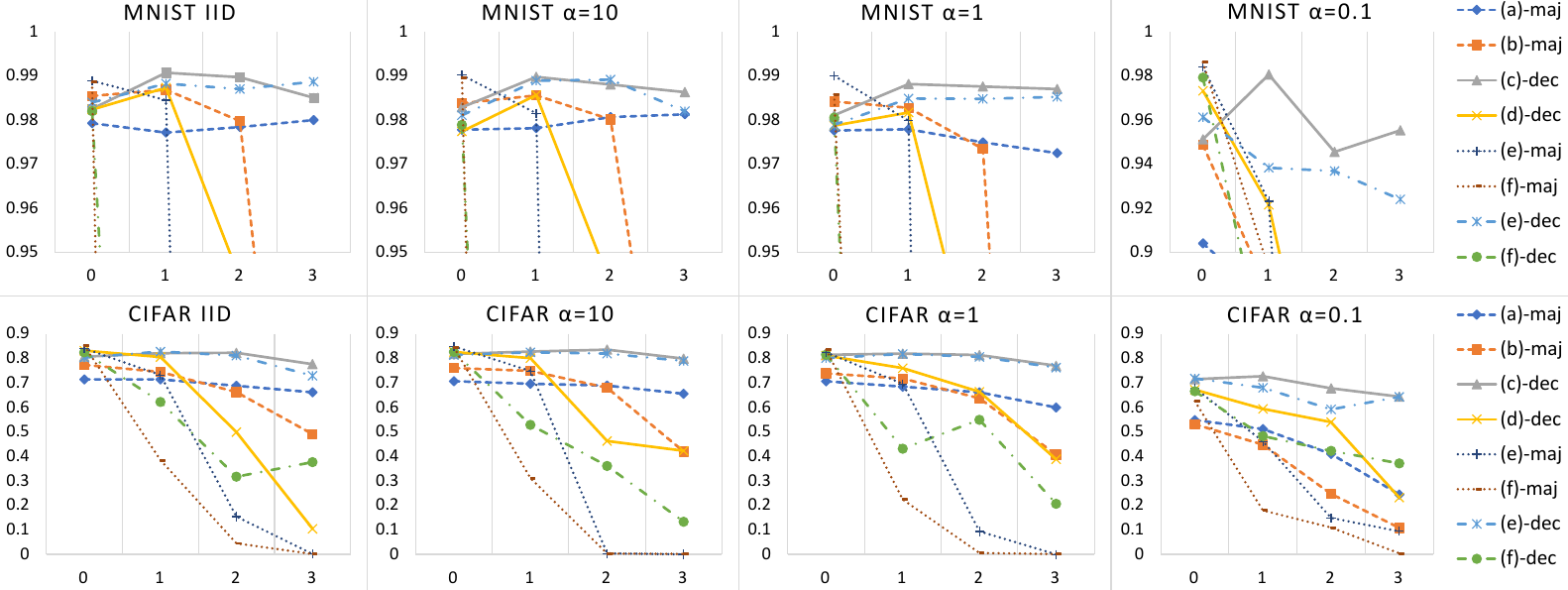}
    \caption{Classification accuracy of BCC codes under 0$\sim$3 attackers.}
    \label{fig:expriments}
\end{figure*}

\subsection{Backdoor Correction}
We use classification accuracy under attack to evaluate the backdoor correction performance.
Figure~\ref{fig:expriments} summarizes the results from the two datasets with four different IID/non-IID settings.
Each line represents the classification accuracy (averaged from 5 repeated runs) of a code under 0$\sim$3 attackers (0 means no attack).
Code (a) and (b) are decoded by majority vote.
Code (c) and (d) are decoded by the proposed decoder.
Code (e) and (f) are decoded using both approaches, denoted as (e/f)-\textit{maj} and (e/f)-\textit{dec} respectively.
We have the following observations from the results:
\begin{itemize}
    \item Codes decoded by the proposed decoder have a slight drop in accuracy on clean data, because some clean data are mistakenly considered as attacks.
    Despite the above effect, code (a) and (b) have the lowest clean data accuracy in general due to the low data utilization ratio.
    The accuracy drop is more prominent for smaller $\alpha$.
    \item Codes decoded by majority vote perform poorly when $r>1/(2k+1)$ due to the majority vote bound. When using the propose decoder, random codes (e) and (f) have higher robustness but are weaker than the proposed codes. The gap is especially large when data utilization ratio is high (code (d) versus code(f)-dec).
    \item The proposed codes have good performance when the number of attackers is below the codes' designed capability. As code (c) with $k=4$ is effective against 1$\sim$3 attackers, code (d) with $k=2$ can only reliably defend one attacker.
\end{itemize}

\subsection{Backdoor Tracking}
% \textbf{Attacker tracking.}
% Here we use 16 users since the code construction is relatively easy for the power of 2's.
To evaluate the backdoor tracking capability, we use the decoder to predict the binary vector $\mathbf{x}$ that corresponds to the attackers' locations, and compare the predicted attackers with the true attackers.
The metric is based on true positives (TP) and false positives (FP), where TP should be as close to $k$ as possible and FP should be near to zero.
Our decoder is allowed to predict any number of attackers, but it will prioritize on finding the minimum number of attackers when $Q(\|\mathbf{x}\|_1)$ is a uniform distribution.

We perform the experiments using IID data and summarize the results in Table~\ref{tab:tracking}.
Means and standard deviations are gathered from 10 repeated runs.
As we can observe from the results, code (g) can reliably track one attacker, while code (h) can track up to two attackers with a lower accuracy.
Such result is consistent with the codes' theoretical properties.
For codes (a)$\sim$(f), we obtain an average TP of 0.2, 0.5, 0.5 for 1$\sim$3 attackers and an average FP of 0.8, 1.1, 1.2.
The backdoor tracking capability is close to random guess.

% \begin{table*}[]
% \centering
% \footnotesize
% \caption{True-positives and false-positives of attacker tracking.}
% \label{tab:tracking}
% \begin{tabular}{@{}llllllll@{}}
% \toprule
% \multirow{2}{*}{Dataset} & \multirow{2}{*}{$k$} & \multicolumn{2}{l}{Code (a)$\sim$(f) Average} & \multicolumn{2}{l}{Code (g)} & \multicolumn{2}{l}{Code (h)} \\ \cmidrule(l){3-8} 
%  &  & TP & FP & TP & FP & TP & FP \\ \midrule
% \multirow{3}{*}{\begin{tabular}[c]{@{}l@{}}MNIST\end{tabular}} & 1 & 0.26$\pm$0.11 &  0.74$\pm$0.11 & \textbf{1.0}$\pm$0.0 & 0.0$\pm$0.0 & \textbf{1.0}$\pm$0.0 & 0.01$\pm$0.0 \\
%  & 2 & 0.52$\pm$0.23 & 1.12$\pm$0.26 &0.29$\pm$0.42 & 0.71$\pm$0.42 & \textbf{1.99}$\pm$0.0 & 0.01$\pm$0.0  \\
%  & 3 & 0.52$\pm$0.19 & 1.18$\pm$0.45 &0.27$\pm$0.39 & 0.73$\pm$0.39 & 2.68$\pm$0.94 & 0.31$\pm$0.94 \\ \midrule
% \multirow{3}{*}{\begin{tabular}[c]{@{}l@{}}CIFAR\end{tabular}} & 1 & 0.16$\pm$0.09 & 0.89$\pm$0.11 & \textbf{0.97}$\pm$0.02 & 0.03$\pm$0.02 & \textbf{0.83}$\pm$0.21 & 0.38$\pm$0.36  \\
%  & 2 & 0.57$\pm$0.13 & 1.06$\pm$0.25 & 0.18$\pm$0.24 & 0.82$\pm$0.24 & \textbf{1.72}$\pm$0.36 & 0.26$\pm$0.27  \\
%  & 3 & 0.44$\pm$0.34 & 1.16$\pm$0.27 & 0.2$\pm$0.4 & 0.8$\pm$0.4 & 2.36$\pm$0.56 & 0.36$\pm$0.34  \\ \bottomrule
% \end{tabular}
% \end{table*}

\begin{table}[]
\centering
\footnotesize
\caption{True-positives and false-positives of attacker tracking.}
\label{tab:tracking}
\begin{tabular}{@{}llllll@{}}
\toprule
\multirow{2}{*}{Dataset} & \multirow{2}{*}{$k$} & \multicolumn{2}{l}{Code (g)} & \multicolumn{2}{l}{Code (h)} \\ \cmidrule(l){3-6} 
 &  & TP & FP & TP & FP \\ \midrule
\multirow{3}{*}{\begin{tabular}[c]{@{}l@{}}MNIST\\IID\end{tabular}} & 1 & \textbf{1.0}$\pm$0.0 & 0.0$\pm$0.0 & \textbf{1.0}$\pm$0.0 & 0.01$\pm$0.0 \\
 & 2 &0.29$\pm$0.42 & 0.71$\pm$0.42 & \textbf{1.99}$\pm$0.0 & 0.01$\pm$0.0  \\
 & 3 &0.27$\pm$0.39 & 0.73$\pm$0.39 & 2.68$\pm$0.94 & 0.31$\pm$0.94 \\ \midrule
\multirow{3}{*}{\begin{tabular}[c]{@{}l@{}}CIFAR\\IID\end{tabular}} & 1 & \textbf{0.97}$\pm$0.02 & 0.03$\pm$0.02 & \textbf{0.83}$\pm$0.21 & 0.38$\pm$0.36  \\
 & 2 & 0.18$\pm$0.24 & 0.82$\pm$0.24 & \textbf{1.72}$\pm$0.36 & 0.26$\pm$0.27  \\
 & 3 & 0.2$\pm$0.4 & 0.8$\pm$0.4 & 2.36$\pm$0.56 & 0.36$\pm$0.34  \\ \bottomrule
\end{tabular}
\end{table}

\subsection{Discussion}
% In general, for a given attack setting, the code should have its $k$ be at least equal to the number of attackers, and higher when the model is noisy (having a low clean data accuracy).
% When the minimal $k$ is satisfied, the code should have its $r$ as high as possible.
% The total cost $m$ is linear to $k$ at minimum but could grow exponentially fast when a high $r$ is required.
The above experiments show that our codes achieve their design goal, which is to obtain a high data utilization ratio while maintaining the robustness.
A significant gain in classification accuracy can be observed on non-IID datasets.
Comparison with random matrices indicates the importance of careful code design, especially when data utilization ratio is high.

Improving the data utilization or tracking the attackers comes with the cost of using more models.
This limits our method to be mostly effective against a small number of attackers.
However, the scalability issue is a fundamental limitation of all subset aggregation-based method and cannot be addressed by designing better codes.
To reduce the overall cost, potential solutions could be using a large number of cheaper models or sharing parts of computation across different models.
% Another direction is to search for nonlinear codes that cannot be represented by a single matrix.

\section{Related Works}

\subsection{Backdoor Attacks and Defenses}
Backdoor attacks were originated as a data-poisoning attack, where attackers inject backdoors by poisoning a portion of training data~\cite{gu2017badnets, turner2018clean}.
In Federated learning, attackers could directly poison the gradients which further increase the flexibility of the attack~\cite{bagdasaryan2018backdoor}.
Most backdoor defenses rely on extracting abnormal statistical features of backdoored data, which are mostly heuristic based~\cite{tran2018spectral, chen2018detecting}.
When attackers have knowledge of the features used by the defender. they can adversarially create backdoor samples that bypass the defense~\cite{tan2019bypassing}.
% For defenses that reverse engineer backdoor triggers from a pre-trained model~\cite{wang2019neural,qiao2019defending}, attackers may design dynamic backdoors that use different triggers on different inputs~\cite{salem2020dynamic}.
% Conventional backdoor defenses focus on removing poisoned data~\cite{tran2018spectral}, which is ineffective against fake gradients.
% Sun \textit{et al.} show that adding norm bounds and Gaussian noises can mitigate the attack to a certain degree but provide no robustness guarantee~\cite{sun2019can}.
% Another line of works provides post-training defense by rejecting backdoored test inputs or removing backdoors from a pre-trained model~\cite{wang2019neural,qiao2019defending}.
% All these methods requires strong assumption of the backdoor's form (shape and size of the backdoor trigger, etc.).

Provable defenses use ensemble models to guarantee a certain degree of backdoor robustness.
Although ensemble models are costly to train and inference, this is the only know type of defense that can provide robustness regardless of the exact attack method.
In randomized smoothing-based methods~\cite{weber2020rab,wang2020certifying}, the defense assumes that attackers only apply small perturbations (with bounded $L_2$ norm) to the training image.
In subset aggregation-based method~\cite{levine2020DeepPA,jia2020intrinsic}, the defense assumes that attackers only change a small number of training data.
Both types of methods are not developed in the context of collaborative learning and assume that each training sample can be independently poisoned.
In such settings where $n\sim10^4$ (each training sample is considered as a user) and $k\sim10^2$ (hundreds of poisoned samples), our method will have similar solutions as the previous works.

\subsection{Coding Theory and ML}
% Coding theory studies codes and their applications.
% A major junction between coding theory and ML is in the area of error correction codes (ECCs), a family of codes that control errors in data and are used in channel coding and data storage~\cite{richardson2008modern}.
A major junction between coding theory and ML is in the area of error correction codes (ECCs).
% Most ECCs are linear codes that can be formulated as a linear algebra problem that solves a sparse error vector $\mathbf{e}$ given a syndrome $\mathbf{y}=H(G\mathbf{x}+\mathbf{e})=H\mathbf{e}$.
% Here $\mathbf{x}$ is a message, $G$ is a generator matrix, and $H$ is a parity-check matrix that always has $H(G\mathbf{x})=0$.
% For binary codes, the algebra is defined on $\mathbb{Z}_2$ with AND as multiplication and XOR as addition.
ECCs are applied to ML most notably as error correcting output codes~\cite{dietterich1994solving} for improving model robustness~\cite{verma2019error}.
In Federated learning, DRACO~\cite{chen2018draco} uses ECCs for gradient encoding and provide Byzantine resilience by introducing redundant gradients.
% Conversely, ML is also applied to design more efficient channel codes~\cite{kim2018deepcode} and more robust decoders~\cite{jiang2019deepturbo}.

Superimposed codes emerge from information retrieval~\cite{kautz1964nonrandom} and have applications in non-adaptive group testing~\cite{du2000combinatorial}.
Several variants of the code such as disjunct matrices and separable matrices are discussed in~\cite{chen2007exploring}.
%: the Boolean sum of any $d$ columns of the matrix does not contain any other column outside of the $d$ columns in the sum ($A \text{ contains } B \Leftrightarrow A\lor B=A$)~\cite{chen2007exploring}.
% A $\bar{d}$-separable matrix requires the Boolean sum of any $\le d$ columns of the matrix to be unique.
% For a code matrix $H$, let $\{\mathbf{h}_1,\dots,\mathbf{h}_k\}$ be any $1<k\le d+1$ different columns of $H$, then $\bigvee_{i=1}^k\mathbf{h}_i\neq\bigvee_{i=1}^{k-1}\mathbf{h}_i$.
% Superimposed codes have a strong relationship with our Backdoor Tracking Codes and provide a lower bound the optimal code design.
\section{Conclusion}

We propose a new subset aggregation-based defense against backdoor attacks in collaborative learning and overcome the data utilization issue in previous defenses.
A coding-based framework is introduced to analyze the optimal subset selection.
Theoretically, we prove the fundamental tradeoff between data utilization ratio and backdoor robustness, which indicates the limitation of all subset aggregation methods.
In practice, we show that by combining optimal codes with a probabilistic decoder, our defense can achieve a high classification accuracy while maintaining backdoor robustness.

\bibliography{reference}

\section*{Proofs and Derivations}

% \subsection{Defending a single attacker}

% \textbf{Proposition.}
% If $H$ is BTC$(1,1,n)$, then $H$ has its number of rows $m\ge\ceil{\log(n+1)}+1$.

% \textit{Proof.}
% For a given $m$, denote $S$ as the set of all binary sequences with length $m$. Then $S$ includes $2^m$ elements.
% For $a\in S$ and $b\in S$, denote $a\sim b$ if and only if the XOR of $a$ and $b$ equals $\mathbf{1}$.
% It can be verified that $\sim$ defines a equivalence relation and $S/\sim$ has $2^{m-1}$ elements.
% If $H$ is BTC$(1,1,n)$, then any two columns $a$ and $b$ in $H$ have $a\not\sim b$.
% Therefore if $H$ has $m$ rows, $n\le 2^{m-1}$.
% Additionally, column in $H$ cannot be $\mathbf{0}$, therefore $n\le 2^{m-1}-1$.
% For integers $m$ and $n$, rewriting the inequality $n\le 2^{m-1}-1$ gives $m\ge\ceil{\log(n+1)}+1$.
% \hfill\qedsymbol

% \subsection{Defending multiple attackers}

\textbf{Theorem 3.}
For $k\ge1$, $r\ge1$, if $H$ is BDC$(k,r,k+r)$, then its number of rows $m\ge\binom{k+r}{k}$.
The solution for $m=\binom{k+r}{k}$, denoted as $\mathcal{H}^{(k,r)}$, is unique up to row and column permutations and can be constructed by:
\begin{gather*}
\mathcal{H}^{(k,1)}=I_{k+1},\\
\mathcal{H}^{(1,r>1)}=\begin{bmatrix}
    \mathbf{1} & \mathcal{H}^{(1,r-1)}\\
    0 & \mathbf{1}^T\end{bmatrix},\\
\mathcal{H}^{(k>1,r>1)}=\begin{bmatrix}
    \mathbf{1} & \mathcal{H}^{(k,r-1)}\\
    \mathbf{0} & \mathcal{H}^{(k-1,r)}\end{bmatrix}.
\end{gather*}

\textit{Proof.}
% Step 1.
We first prove the lower bound of $m$, which includes three cases:

Case 1: rows are unique and all rows have exactly $k$ 0's.
Assume $H$ is BDC$(k,r,k+r)$ with $m<\binom{k+r}{k}$.
Since rows of $H$ are unique and there exists $\binom{k+r}{k}$ different binary sequences with $k$ 0's and $r$ 1's, at least one binary sequence with $k$ 0's and $r$ 1's does not appear in any row of $H$.
Denote this sequence as $a$ and the indices of 0's in $a$ as $S=\{i_1, i_2, \dots, i_k\}$.
Select elements of $a$ by the index set $S$ and denote the result as $a_S=\{a_{i_1}, a_{i_2}, \dots, a_{i_k}\}$, then all elements of $a_S$ are 0's.
For any other binary sequence $b$ with $k$ 0's and $r$ 1's such that $b\neq a$, select elements of $b$ using the same index set $S$.
The result $b_S$ has at least one 1 and the Boolean sum of $b_S$ is 1.
Let $H_S$ be the $k$ columns of $H$ selected by $S$, and $s$ be the Boolean sum of $H_S$, then $s$ is a $k$-sum of $H$.
Since $a$ is not in $H$, we have $s=\mathbf{1}$, which contradicts the assumption of $H$ being BDC$(k,r,k+r)$.

Case 2: rows are unique and some rows have less than $k$ 0's.
If any row has less than $k$ 0's, then any $k$-sum of $H$ equals 1 in this row.
If $H$ is BDC$(k,r,k+r)$ then the code obtained by removing this row from $H$ is also BDC$(k,r,k+r)$.
Repeat the process until all rows with less than $k$ 0's are removed, then the problem is reduced to Case 1.

Case 3: rows are not unique.
Similarly, if $H$ is BDC$(k,r,k+r)$ then the code obtained by removing a non-unique row from $H$ is BDC$(k,r,k+r)$.
Repeat the process until all non-unique rows are removed, then the problem is reduced to Case 2.
Since there are no other possible cases, $H$ always has $m\ge\binom{k+r}{k}$.

% Step 2.
Next we verify that $\mathcal{H}^{(k,r)}$ has $m=\binom{k+r}{k}$ and satisfies the requirements of BDC$(k,r,k+r)$.

When $r=1$, obviously $m=k+1$ and $I_{k+1}$ is BDC$(k,1,k+1)$.

When $k=1$ and $r>1$, assume that $\mathcal{H}^{(1,r-1)}$ is BDC$(1,r-1,r)$ and has $m=r$. Then $\mathcal{H}^{(1,r)}$ has $m=r+1$.
Consider the row weight of the first $r-1$ rows, the result is 1 plus the row weight of $\mathcal{H}^{(1,r-1)}$, which gives $1+(r-1)=r$.
Since the last row of $\mathcal{H}^{(1,r)}$ has $r$ 1's, $\mathcal{H}^{(1,r)}$ has a row weight of $\min\{r,r\}=r$.
The last $r-1$ columns are not $\mathbf{1}$ since their first $r-1$ element are not all 1's (based on the assumption of $\mathcal{H}^{(1,r-1)}$).
The first column is not $\mathbf{1}$ by construction.
Since $\mathcal{H}^{(1,r)}$ has a row weight of $r$ and no column equals $\mathbf{1}$, $\mathcal{H}^{(1,r)}$ is BDC$(1,r,r+1)$.

When $k>1$ and $r>1$, assume that $\mathcal{H}^{(k,r-1)}$ has $m=\binom{k+r-1}{k}$ and $\mathcal{H}^{(k-1,r)}$ has $m=\binom{k+r-1}{k-1}$.
Then $\mathcal{H}^{(k,r)}$ has $m=\binom{k+r-1}{k}+\binom{k+r-1}{k-1}=\binom{k+r}{k}$.
Further assume that $\mathcal{H}^{(k,r-1)}$ is BDC$(k,r-1,k+r-1)$ and $\mathcal{H}^{(k-1,r)}$ is BDC$(k-1,r,k+r-1)$.
Then for $\mathcal{H}^{(k,r)}$, the row weight is given by the minimum of the row weights of the first $\binom{k+r-1}{k}$ rows and the last $\binom{k+r-1}{k-1}$ rows, which is $\min\{1+(r-1),r\}=r$.
For any $\bar{k}$-sum in $\mathcal{H}^{(k,r)}$, if the addends of the $\bar{k}$-sum includes the first column, then the $\bar{k}$-sum is not $\mathbf{1}$ since the last $\binom{k+r-1}{k-1}$ elements of the $\bar{k}$-sum is equal to a $\overline{k-1}$-sum of $\mathcal{H}^{(k-1,r)}$, which is not $\mathbf{1}$.
If the addends of the $\bar{k}$-sum does not include the first column, then the $\bar{k}$-sum is not $\mathbf{1}$ since the first $\binom{k+r-1}{k}$ elements of the $\bar{k}$-sum is equal to a $\bar{k}$-sum of $\mathcal{H}^{(k,r-1)}$, which is not $\mathbf{1}$.
Since $\mathcal{H}^{(k,r)}$ has a row weight of $r$ and no $\bar{k}$-sum equals $\mathbf{1}$, $\mathcal{H}^{(k,r)}$ is BDC$(k,r,k+r)$.

By induction, $\mathcal{H}^{(k,r)}$ is BDC$(k,r,k+r)$ and has $m=\binom{k+r}{k}$ for all positive integers $k$ and $r$.

% Step 3.
At last we show that the solution for $m=\binom{k+r}{k}$ is unique up to row and column permutations.
As proved earlier, if $m=\binom{k+r}{k}$, then all rows are unique and each row includes exactly $k$ 0's.
Since there are only $\binom{k+r}{k}$ possible binary sequences with $k$ 0's and $r$ 1's, all of these sequences must be selected exactly once to construct the code, which implies uniqueness of the selection. %being unique.
\hfill\qedsymbol

\textbf{Theorem 5.}
For $k\ge1$, $r>1$, the minimal BCC$(k,r,k+r)$ has $m=\binom{k+r}{k}$.
For $r=1$, the minimal BCC$(k,1,k+1)$ has $m=k+2$.

\textit{Proof.}
For $k\ge1$, $r>1$, Theorem~3 states that $\mathcal{H}^{(k,r)}$ with $m=\binom{k+r}{k}$ is BDC$(k,r,k+r)$.
We need to verify that $\mathcal{H}^{(k,r)}$ is also BCC$(k,r,k+r)$, meaning that the XOR of any two $\bar{k}$-sums is not equal to $\mathbf{1}$.

When $k=1$, $r>1$, every columns of $\mathcal{H}^{(1,r)}$ have more 1's than 0's.
By monotonicity of Boolean sums, every $\bar{k}$-sums of $\mathcal{H}^{(1,r)}$ have more 1's than 0's.
If the XOR of two $\bar{k}$-sums equals $\mathbf{1}$, their total number of 1's must be equal to their total number of 0's, which is impossible.

When $k>1$, $r>1$, assume that the XOR of any two $\bar{k}$-sums is not equal to $\mathbf{1}$ for $\mathcal{H}^{(k-1,r)}$, then consider $\mathcal{H}^{(k,r)}$ in three cases:

Case 1: no $\bar{k}$-sum of $\mathcal{H}^{(k,r)}$ includes the first column.
The XOR of any two $\bar{k}$-sums is not equal to $\mathbf{1}$ since the last $\binom{k+r-1}{k-1}$ elements are not all 1's (guaranteed by $\mathcal{H}^{(k-1,r)}$).

Case 2: one and only one $\bar{k}$-sum of $\mathcal{H}^{(k,r)}$ includes the first column. The first $\binom{k+r-1}{k}$ elements of one $\bar{k}$-sum are all 1's.
The first $\binom{k+r-1}{k}$ elements of the other $\bar{k}$-sum have at least one 1 since $\mathcal{H}^{(k,r-1)}$ contains no zero column.
The XOR of the two $\bar{k}$-sums has at least one 0.

Case 3: both $\bar{k}$-sums of $\mathcal{H}^{(k,r)}$ include the first column. Obviously the first $\binom{k+r-1}{k}$ elements of the XOR of the two $\bar{k}$-sums are equal to 0.

By induction, $\mathcal{H}^{(k,r)}$ is BCC$(k,r,k+r)$ for all positive integers $k$ and $r>1$.

For $r=1$, obviously $I_{k+1}$ is BDC but not BCC. With Lemma~4, adding an additional row of 1's to $I_{k+1}$ turns the code to be BCC$(k,1,k+1)$, which gives $m=k+2$.
\hfill\qedsymbol

\subsection{Probabilistic Decoding}

For model $i$, we use confusion matrices $C^{(i)}_{jk}$ to estimate the probability of classifying clean data in class $j$ as class $k$, which gives:
\[
Pr(\text{pred}_i=y_i|\text{attack}=\text{False},\text{label}=l)=C_{ly_i}^{(i)},
\]
in which \textit{$\text{pred}_i=y_i$} is the class prediction of model $i$, \textit{$\text{attack}=\text{False}$} refers to the cases that no attack happens, and \textit{$\text{label}=l$} represents the true label of the data.

Assuming that the attack has a fixed target \textit{$\text{targ}=t$}, then when an attack succeeds, denoted as \textit{$\text{atkSucc}=\text{True}$} (do not confuse with \textit{$\text{attack}=\text{True}$}, which means an attack happens but not necessarily succeeds), we have:
\begin{gather*}
Pr(\text{pred}_i=y_i|\text{atkSucc}=\text{True},\text{targ}=t)=\delta_{ty_i} 
\\
\text{in which}\quad
\delta_{ij}=\begin{cases}1\text{ if }i=j\\0\text{ otherwise}\end{cases}.
\end{gather*}
Further assuming that a failed attack does not affect the model's prediction, which gives:
\[
Pr(\text{pred}_i=y_i|\text{atkSucc}=\text{False},\text{label}=l)=C_{ly_i}^{(i)}.
\]
The rate of attack $A$ specifies our prior assumption of the likelihood of an attack to happen:
\[
Pr(\text{attack}=\text{True})=A.
\]
The success rate of attack $S$ connects the events of an attack to happen and to succeed as:
\begin{gather*}
Pr(\text{atkSucc}=\text{True}|\text{attack}=\text{True},\text{attackers}=\mathbf{x})\\
=\begin{cases}S &\text{if }\sum_j H_{ij}x_j>0\\0 &\text{otherwise}\end{cases}.
\end{gather*}
Here $H$ is the code matrix, $\mathbf{x}$ is the binary vector representing all attackers, and $H_{ij}x_j>0$ means that model $i$ is backdoored.

Connecting the above equations gives:
\begin{gather*}
Pr(\text{pred}_i=y_i|\text{attack}=\text{True},\text{attackers}=\mathbf{x},\text{targ}=t,\\\text{label}=l)\\
=\begin{cases}S\delta_{ty_i}+(1-S)C^{(i)}_{ly_i}&\text{if }\sum_j H_{ij}x_j>0\\C^{(i)}_{ly_i} &\text{otherwise}\end{cases}.
\end{gather*}

To further specify the distribution of the locations of the attackers, we use $Q(k)$ to represent the probability of having $k$ attackers out of $n$ users.
The permutation symmetry of the locations gives:
\[
Pr(\text{attackers}=\mathbf{x})=\frac{Q(\|\mathbf{x}\|_1)}{\binom{n}{\|\mathbf{x}\|_1}}.
\]

Assuming independent model predictions, we denote \textit{$\text{pred}=\mathbf{y}$} as the class predictions of all models and define $p(\mathbf{x},\mathbf{y},t,l)$ as:
\begin{gather*}
p(\mathbf{x},\mathbf{y},t,l)\vcentcolon=
Pr(\text{pred}=\mathbf{y},\text{attackers}=\mathbf{x}|\text{attack}=\text{True},\\ \text{targ}=t,\text{label}=l)\\
=\frac{Q(\|\mathbf{x}\|_1)}{\binom{n}{\|\mathbf{x}\|_1}}\prod_i\begin{cases}S\delta_{ty_i}+(1-S)C^{(i)}_{ly_i}&\text{if }\sum_j H_{ij}x_j>0\\C^{(i)}_{ly_i} &\text{otherwise}\end{cases}.
\end{gather*}

Then assuming attack targets and true labels are independent and uniformly distributed in $c$ classes, i.e., $Pr(\text{targ}=t,\text{label}=l)=1/c^2$, we have:
\begin{gather*}
Pr(\text{pred}=\mathbf{y}|\text{attack}=\text{True})=\sum_{\mathbf{x},t,l} p(\mathbf{x},\mathbf{y},t,l)/c^2
\text{, and}\\
Pr(\text{pred}=\mathbf{y}|\text{attack}=\text{False})=\sum_l\prod_i C^{(i)}_{ly_i}/c.
\end{gather*}

Using Bayes's rule, we finally arrive at:
\begin{gather*}
Pr(\text{attack}=\text{True}|\text{pred}=\mathbf{y})=\\
\frac{A\sum_{\mathbf{x},t,l} p(\mathbf{x},\mathbf{y},t,l)}{A\sum_{\mathbf{x},t,l} p(\mathbf{x},\mathbf{y},t,l)+(1-A)c\sum_l\prod_i C^{(i)}_{ly_i}}.
\end{gather*}

Let $m$ be the number of models, the time complexity of the decoder is $O(n^kmc^2)$.
\section*{Experimental Details}

\subsection{Data Preparation}
\textbf{MNIST.} The dataset contains 60,000 images in the training set and 10,000 images in the testing set. All images in the dataset are normalized before training. The dataset can be downloaded from \url{http://yann.lecun.com/exdb/mnist/}.

\noindent\textbf{CIFAR-10.} The dataset contains 50,000 images in the training set and 10,000 images in the testing set. Each image is randomly cropped into 32px by 32px, horizontally flipped and normalized before training. The dataset can be downloaded from \url{https://www.cs.toronto.edu/~kriz/cifar.html}.

All experiments are performed on 12 or 16 users.
In each experiment, we randomly permute the training set and split the training dataset according to the Dirichlet distribution.
The testing set is split to 2 parts.
Among them, 5,000 images are used for confusion matrix estimation.
The other 5,000 images are used for testing the decoder's defense capability.

\subsection{Model Design}
\textbf{CNN for MNIST.}
The network consists of 2 convolutional layers and 2 fully connected layers.
The first convolutional layer has 32 3\x3 convolution kernels with a stride of 1 and ReLU activation.
The second convolutional layer has 64 3\x3 convolution kernels with a stride of 1, followed by a 2\x2 max-pooling layer and a dropout layer. 
Followed by the convolutional layers are two fully connected layers with output size 128 and 10.

\noindent\textbf{ResNet18 for CIFAR.}
The ResNet18 is adopted from an open source implementation \url{https://github.com/kuangliu/pytorch-cifar}.

\subsection{Training}
We use Adam optimizer of model training with a batchsize of 128.
When a model is trained on data from multiple users, we iterate over the selected users with the granularity of a batch.
Data in a single batch always come from a single user.

For hyper-parameter tuning, we manually explore the poison ratio from 1e-2 to 1e-1, the learning rate from 1e-4 to 1e-2, and the number of training epochs from 10 to 30.
The final results are obtained using the following settings:
The attack uses randomly generated 3px\x3px triggers with 5\% poison ratio. The triggers are generated by a random permutation of five 1's and four 0's on MNIST and by 27 i.i.d. random RGB values following a clipped Gaussian distribution on CIFAR. 
We choose Adam as the optimizer, with learning rate 1e-3 and weight decay 1e-4. Models are trained for 10 epochs on MNIST and 20 epochs on CIFAR. 
The experiment are done on a server with two Intel Xeon E5-2687W CPUs, 755 GiB of memory, and four Nvidia TITAN RTX GPUs.

\subsection{Testing}
In the testing stage, we fix the attack rate $A$ to 0.5 and the attack success rate to 0.99. The attacker distribution $Q$ is a uniform distribution covering 0$\sim$3 attackers.

\end{document}